\documentstyle[epsfig]{aipproc}
\newcommand{\lba}{\lambda}
\newcommand{\lbar}{\lba{\!\!\!\mathchar'26}}
\newcommand{\hba}{h}
\newcommand{\hq}{\hba{\!\!\!\!\mathchar'26\,}}
\begin{document}
\title{Muonium}
\author{Klaus P. Jungmann}
\address{Physikalisches Institut, Universit\"at Heidelberg\\
Philosophenweg 12, D-69129 Heidelberg, Germany}
\maketitle
\begin{abstract}
The energy levels of the muonium ($\mu^+ e^-$) atom, which consists of two
''point-like'' leptonic particles, can be calculated to
very high accuracy in the framework of bound state Quantum Electrodynamics
(QED),
since there are no complications due to internal nuclear structure and size
which is the case for all atoms and ions of natural isotopes including
atomic hydrogen. The ground state hyperfine splitting
and  the 1s-2s energy interval can provide both tests of QED theory
and determinations of fundamental constants like the muon mass $m_{\mu}$
and the fine structure constant $\alpha$.
The excellent understanding of the electromagnetic binding in the
muonium  atom allows in particular testing fundamental physical
laws like lepton number conservation in searches for muonium to
antimuonium conversion and probing the nature of the muon as a
heavy leptonic object.
\end{abstract}

\section{Introduction}

Atomic hydrogen has played an important role in the history of modern
physics.
The successful description of the
spectral lines of the hydrogen atom
by the Schr\"odinger equation \cite{Schro_26}
and especially by the Dirac equation \cite{Dar_28}
had a large
impact on the development of quantum mechanics.
Precision measurements of the ground state hyperfine structure splitting
by Nafe, Nelson and Rabi \cite{Naf_47,Naf_48} in the late 1940's were important
contributions to the
identification of
the magnetic anomaly of the electron.
Together with the observation of the ``classical''
$2^2S_{1/2}-2^2P_{1/2}$ Lamb shift by Lamb and Retherford \cite{Lam_50}
they pushed the development of
the modern theory of Quantum
Electrodynamics(QED).

Today the anomalous magnetic moment of the electron, which is defined as
$a_e=(g_e-2)/2$,
where $g_e$ is the electron g-factor, and the
ground state hyperfine structure splitting $\Delta \nu_{HFS}^H$
of atomic hydrogen are among
the most well
known quantities in physics. Experiments in Penning traps
with single electrons \cite{VDy_90}
have determined $a_e$=1\,159\,652\,188.4(4.3)$\cdot10^{-12}$ to
3.7~ppb.
QED calculations
have reached such a precision that the
fine structure constant $\alpha$ can be extracted to 3.8~ppb
\cite{Kin_97}.
Hydrogen hyperfine structure measurements yield $\Delta
\nu^H_{HFS}= 1\,420.405\,751\,766\,7(9)$~MHz (0.006~ppb) \cite{Ess_71,Hel_70}
and hydrogen masers  even have a large potential as frequency standards.
However, the theoretical description of the hyperfine splitting
is limited to the ppm level
by the fact that the proton's internal structure is not known well enough for
calculations of nearly similar accuracy \cite{Sap_90,Yen_92}.
Neither can experiments supply the necessary data on the
mean square charge radius and the polarizability,
for example from  electron scattering, nor is any theory,
for example low energy quantum chromodynamics (QCD), in a position
to yield the protons internal charge structure and the dynamical
behavior of its charge carrying constituents.
The situation is similar for the $2S_{1/2}$- $2P_{1/2}$
Lamb shift in hydrogen, where
atomic interferometer experiments find
$\Delta \nu_{2S_{1/2}-2P_{1/2}}^H$= 1\,057.851\,4(19)~MHz \cite{Pal_85}
(1.8~ppm)
and
where the knowledge of the proton's mean square charge radius limits any
calculations to the 10~ppm level of precision \cite{Sap_90,Yen_92}.

The  1s-2s level separation in atomic hydrogen has reached a fascinating
accuracy for optical spectroscopy and one can expect even further significant
improvements. The Rydberg constant has been extracted
to R$_\infty$= 10 973 731.568 639(91) m$^{-1}$ \cite{Beau_97,Udem_97}
by comparing this
transition with others in hydrogen and is now the best known fundamental
constant. However, the knowledge of the mean square charge radius limits
comparison between experiment and theory. Detailed reviews on this topic
were given by  \cite{Hansch_98,Pachucki_98} on this conference.

\begin{table}[thb] \centering
 \caption[Hyperfine structure and 1S-2S splitting in some hydrogen-like systems]
    {\label{HYDATOM} {\it
    The ground state hyperfine structure splitting $\Delta \nu_{HFS}$
    and the 1s-2s level separation $\Delta \nu_{1S-2S}$
    of hydrogen and  some
    exotic hydrogen-like systems offer narrow transitions for
    studying the interactions in Coulomb bound two-body systems.
    In the exotic systems the linewidth has a fundamental lower limit
    given by the finite lifetime of
    the systems, because of annihilation, as in the case of positronium,
    or because of weak muon or pion decay.
    The very high
    quality factors (transition frequency divided by the natural linewidth
    $\Delta \nu / \delta \nu$ ) in
    hydrogen and other systems with hadronic nuclei can hardly
    be utilized to test
    the theory because of the insufficiently known charge distribution and
    dynamics of the charge carrying constituents within the hadrons.
    }}
 \vspace{2mm}
 {\small
 \begin{tabular}[t]{|c|*{6}{r@{}l|}}
\hline
 &&{}    &&{}    &&{}    &&{}    &&{}    &&{}\\
 &Posit&{ronium}  &Muon&{ium}   &Hydr&{ogen}  &Muo&{nic}
 &Pion&{ium}        &Muo&{nic} \\
 &&  &&   &&  &Heliu&{m4}
 &&         &Hydr&{ogen} \\
%
 &{\large $e^+$}&{{\large $e^-$}}
 &{\large $\mu^+$}&{{\large $e^-$}}
 &{\large $p^{ }$}&{{\large $e^-$}}
 &{\large $(\alpha\mu^-$}&{{\large $)e^-$}}
 &{\large $\pi^+$}&{{\large $e^-$}}
 &{\large $ p^{ }$}&{{\large $\mu^-$}} \\
 &&{}    &&{}    &&{}    &&{}    &&{}    &&{}\\
\hline
%
  &&{}    &&{}    &&{}    &&{}    &&{}    &&{}\\
{\large{$\Delta\nu _{1S-2S}$}} &123&{3.6}$^{\dagger}$
                                  &245&{5.6}   &246&{6.1}  &246&{8.5}
 &245&{8.6}         &4.59&{$\times 10^5$} \\
$[$THz$]$       &&{}    &&{}    &&{}    &&{}    &&{}    &&{}\\
        &&{}    &&{}    &&{}    &&{}    &&{}    &&{}\\
{\large{$\delta \nu_{1S-2S}$}} &1&{.28}$^{\dagger}$
                                        &.&{145} &1.3&{$\times10^{-6}$} &.&{145}
 &12.&{2}   &.&{176}\\
$[$MHz$]$       &&{}    &&{}    &&{}    &&{}    &&{}    &&{}\\
 &&{}    &&{}    &&{}    &&{}    &&{}    &&{}\\
{\large{$\frac{\Delta\nu_{HFS}}{\delta \nu_{HFS}}$}}
  &1.7&{$\times10^2$}    &3.1&{$\times10^4$}    &3.2&{$\times10^{24}$}
  &3.1&{$\times10^4$}    &-&{-}    &3.1&{$\times10^8$}\\
 &&{}    &&{}    &&{}    &&{}    &&{}    &&{}\\
\hline
 \end{tabular}
\hspace{2mm}
$^\dagger${\footnotesize Only the 1S-2S splitting in the triplet
system of positronium is considered in this table.}
}
\end{table}

In addition to the information obtainable from
the natural isotopes hydrogen ($pe^-$),
deuterium ($de^-$) and
tritium ($te^-$)
or hydrogen-like ions of natural elements ($ (^Z_AX e^-)^{(Z-1)+}$),
exotic hydrogen-like atoms can provide further information about
the interactions between the bound particles
and the nature of these objects themselves \cite{Jun_94}.
Such systems can be formed by replacing the electron in a natural atom
by an exotic particle (e.g. $\mu^-,\pi^-,K^-,\overline{p}$),
or by electron capture of an exotic "nucleus" (e.g. $e^+,\mu^+,\pi^+$).
Even atoms consisting of two
exotic particles have been produced occasionally, e.g. $\pi^+ \mu^-$ and $\pi^-
\mu^+$ were found in in-flight decays of neutral kaons $K_L^0$ \cite{Coo_77}.
Some of the systems are compared in Table \ref{HYDATOM} with respect to the
possible spectroscopic resolution for the 1S-2S and the ground state hyperfine
structure transition.\\

Muonic atoms (($\mu^-$$^A_ZX$)) and ions (($\mu^-$$^A_ZX)^{n+}$)
are of particular interest for spectroscopy, since
the Bohr radius $a_{\mu}$ is about 207 times smaller for muons than for
electrons,
$a_{\mu}= (m_e/m_{\mu})a_0$,
where $m_e$ and $m_{\mu}$ are the masses of the electron and the muon and
$a_0= \hq^2/(m_e e^2)= 0.529\cdot10^{-10}$~m, with the electric charge unit $e$
and Planck's constant $\hq$.
Bound muonic states are therefore more sensitive
to the properties of the nuclei. This has been widely
applied for determinations of nuclear charge moments and
for examining nuclear polarization
\cite{Sch_92,Ros_92a}.
The muon orbit for higher nuclear charges $Z$ is significantly smaller
than
the electron Compton wavelength
$\lbar_C = \hq/(m_e c) =
3.86\cdot 10^{-13}$~m, which is
the typical dimension of
vacuum fluctuations.
In contrast to electronic systems,
 the
vacuum polarization contributions
in muonic atoms are substantially larger than the self energy,
since
the Uehling potential, which describes to lowest order
the modification of the nuclear
potential due to vacuum fluctuations,
scales
approximately
with the cube of the particle mass and the
self energy is inversely proportional to it. Higher order vacuum polarization
contributions are needed for a satisfactory
 description.

There is special interest in muonic hydrogen ($p\mu^-$),
since, in principle,  one could obtain from a measurement of its
ground state hyperfine splitting and the 2S-2P Lamb shift
more detailed information on the proton's charge radius
and polarizability \cite{Jun_92a,Bak_93}.

The muonic helium atom ($(\alpha\mu^-)^+e^-$)
\cite{Sou_80,Ort_80,Ort_82}
consists of a pseudo-nucleus $(\alpha\mu^-)^+$,
which is itself a hydrogen-like system,
 and an electron. It is the simplest atomic system
involving both a negative muon and an electron.
A precise value for
the magnetic moment of the negative muon has been deduced
from the Zeeman splitting of the ground state  hyperfine structure
as a test of the CPT theorem.

Already in the 1970's the muonic helium ion $(\alpha\mu^-)^+$
was the first exotic atomic system
for which a successful laser experiment has been reported
\cite{Car_76,Car_77}. Allowed electric dipole transitions between
the n=2 fine structure levels have been induced.
A precise test of QED vacuum polarization could be established
and the rms charge radius of the $\alpha$-particle was extracted.
The experiment needed as a prerequisite the metastability of the 2S state
at high pressures ($\approx$40~bar), which could not be confirmed
in several independent approaches \cite{Arb_84,Eck_86,Ros_90a} and
a second attempt of a laser experiment
could not confirm the existence of a signal \cite{Hau_92}, which
leaves us with an open puzzle.

In purely hadronic systems like pionic hydrogen ($p\pi^-$) and
antiprotonic hydrogen \cite{Bee_91a},
protonium ($p\overline{p}$)
\cite{Aul_78,Kle_89}, the transition frequencies are dominantly due to
Coulomb interaction.
They bear additional line shifts and broadenings
in the spectral lines due to strong interaction between the constituents.
 They offer the possibility to study the strong interaction
between the particles at zero energy.
From the transition frequencies in pionic
atoms (or possibly from the pionium atom ($\pi^+e^-$) \cite{Mun_89}
one can derive a value for the pion mass \cite{Jec_86}.
At present the determination of the best upper
limit for the muon neutrino mass from
the muon momentum from pion decays at rest \cite{Dau_91,Dau_92,assa94}
is spoiled by the
uncertainties
of the pion mass.
\\

For leptons no internal structure is known so far.
Scattering experiments have established that
electron ($e$), muon($\mu$) and tauon ($\tau$) behave like point-like particles
down to dimensions of less than $10^{-18}$m \cite{Mar_90,Kin_90a}
which is three orders
of magnitude below the proton's rms charge radius
\cite{Sic_82,Sim_80,Han_63}.
Purely leptonic hydrogen-like systems like
positronium ($e^+e^-$) \cite{Deu_51,Mil_90}
and
muonium ($\mu^+ e^-$) \cite{Hug_60,Hug_90,Hug_97}
have interesting perspectives for testing bound state QED,
for searching for deviations from present models and for testing
fundamental symmetry laws.

Positronium is a particle anti-particle system and
significantly differs
from muonium and
hydrogen-like systems with different constituent masses.
The Furry picture,
where  the electronic states are in zeroth order
solutions of the Dirac equation in an external electrostatic field
and  which is successfully applied for the heavier systems,
is not appropriate.
A theoretical description must start from the fully relativistic
Bethe-Salpeter
formalism or an approximation to it \cite{Moh_88}.
Depending on the C-parity of the state,
the ground state of the positronium atom annihilates
into two ($1^1S_0$) or three ($1^3S_0$)
photons.
Standard theory was confirmed in
various searches for rare decay modes
which have been carried out \cite{Mil_90}
in order to find unkown light particles, e.g. axions, or violations of
fundamental laws, e.g.  C-parity symmetry.
For the $^3S$-states annihilation into a single virtual photon
causes significant shifts at the fine structure level.
Positronium was the second exotic system in which laser excitation
could be achieved \cite{Chu_84}.
The 1S-2S transition frequency is known to
$\Delta \nu_{1S-2S}^{PS}$ = 1\,233\,607\,216.4(3.2)~MHz (2.6~ppb) corresponding
to a test of the
QED contributions to 35~ppm \cite{Fee_93}.
The theoretical uncertainty is estimated to
be of the order
of 10~MHz from uncalculated higher order ($\alpha^4R_{\infty}$)
terms \cite{Fel_92}.
Their evaluation will be cumbersome, because of the inflation of
the number of Feynman diagrams
due to virtual annihilation.
From the result one can conclude that electron and positron masses are equal to
2~ppb,  the best test of mass equality for particle and antiparticle
next to the $K^0\overline{K^0}$ system \cite{Rev_92},
where the relative mass difference
is $\leq 4 \cdot 10^{-18}$.

Precision experiments on muonium offer a unique opportunity to
investigate bound state QED without complications arising from nuclear structure
and to test the behavior of the muon
as a heavy leptonic particle and hence the electron-muon(-tau) universality,
which is fundamentally assumed in QED theory.
Of particular interest is
the ground state hyperfine structure splitting, where experiment \cite{Mar_82}
and theory \cite{Sap_90,Yen_92}
agree at 300~ppb  which is at a higher level of precision than in
the case of atomic hydrogen. An accurate value for the fine structure constant
$\alpha$ can be extracted to 140~ppb.
The Zeeman effect of the ground state hyperfine sublevels
yields the most precise value for muon the magnetic moment $\mu_{\mu}$
with an accuracy of 360~ppb.
Signals from the "classical" $2^2S_{1/2}$-$2^2P_{1/2}$ Lamb shift in muonium
have been observed \cite {Ora_84,Bad_84}. However,
with 1.4\% precision they are
not yet in a region
where they can be confronted with
theory.
The sensitivity to
QED corrections is highest for the 1S-2S interval in muonium
due to the approximate $1/n^3$ scaling of the Lamb shift.
Compared to hydrogen, the
radiative recoil and the relativistic recoil effects are
larger by a factor of $m_p/m_{\mu} \approx$8.9, where $m_p$ is the proton mass
and $m_{\mu}$ the muon mass.
A hydrogen-muonium isotope shift measurement in this
transition
as well as a technically  only slightly more difficult measurement
of the 1s-2s transition frequency in muonium
can lead to a new and accurate figure for the muon mass $m_{\mu}$.
The transition has recently been excited successfully
in two independent experiments \cite{Chu_88,Dan_89,Jun_91,Maas_94}.

\begin{figure}[hbt]
\unitlength 1cm
\centering
 \begin{picture}(10,10.24)
  \epsfig{file=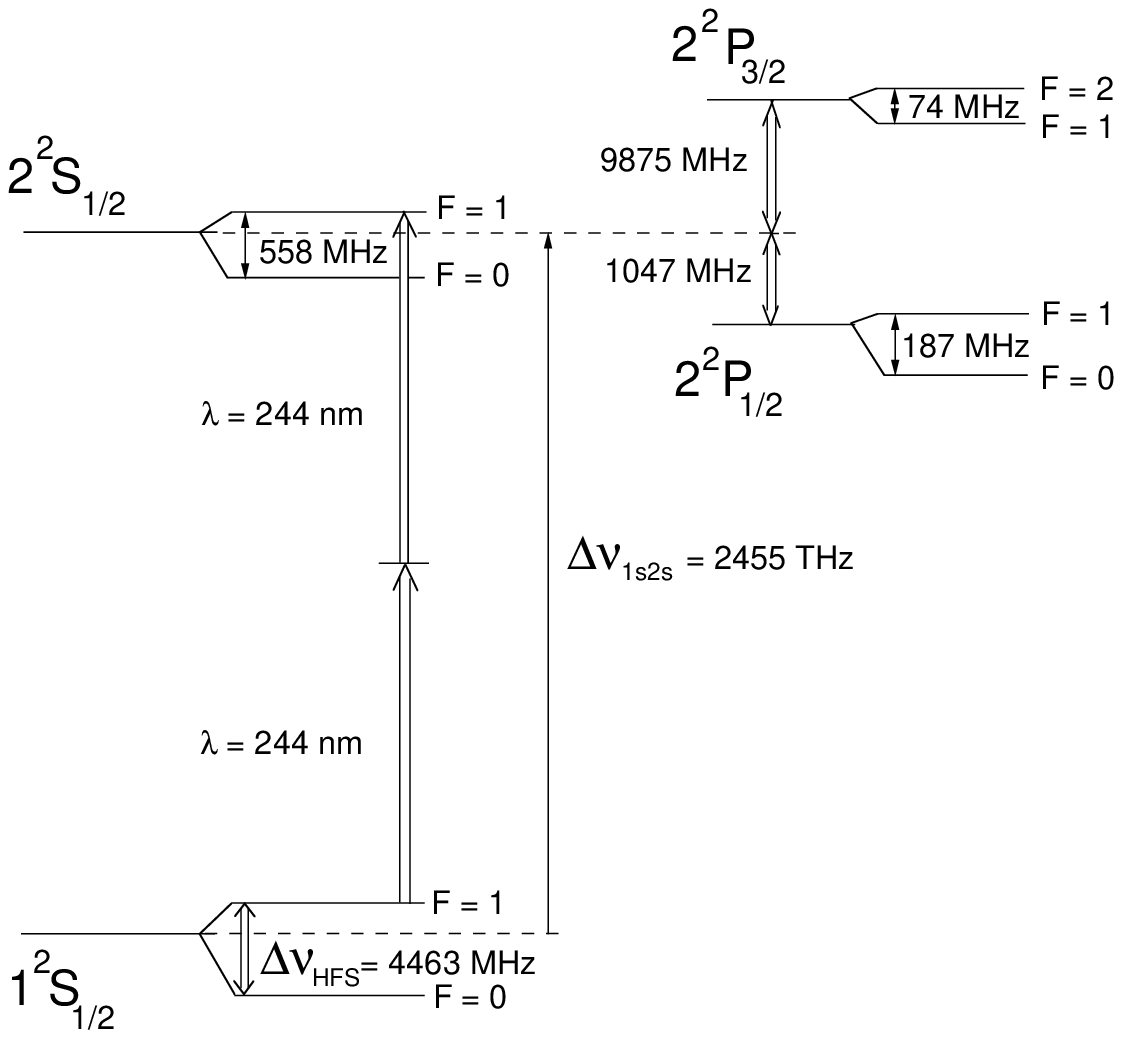,height=10.2cm}
 \end{picture}\par
 \caption[]{ Energy levels of muonium for principal quantum numbers n=1 and n=2.
 The indicated gross, fine and hyperfine structure transistions were studied yet.
 The most accurate measurements are the ones involving
 the n=1 ground state in which the atoms
 can be produced efficiently.}
\end{figure}

The spectroscopic
experiments in muonium are closely inter-related with
the  determination of the muon's
magnetic anomaly $a_{\mu}$ through the relation
$\mu_{\mu}=(1+a_{\mu})~~ e\hq/(2m_{\mu}c)$.
The results from all experiments
establish a self consistency
requirement for QED and electroweak
theory and the set of fundamental constants involved.
The constants $\alpha,m_{\mu},\mu_{\mu}$ are the most stringently tested
important parameters.
The only necessary external input are the hadronic
corrections to $a_{\mu}$ which can be obtained from a measurement of the
ratio of cross sections
$(e^+e^- \rightarrow \mu^+\mu^-)/(e^+e^- \rightarrow hadrons)$.
Although, in principle, the system could provide the relevant electroweak
constants, the Fermi coupling constant $G_F$ and $\sin^2\theta_W$,
the use of more accurate values from
independent measurements
may be chosen for higher sensitivity to new physics.
As a matter of fact,
an improvement upon the present knowledge of the muon mass at the
0.35 ppm level is very important
for the success of a new measurement of the muon magnetic anomaly
presently under way at the Brookhaven National Laboratory in
Upton, New York. The relevance of the experiment, which aims for 0.35 ppm accuracy,
arises from its sensitivity
to contributions from physics beyond the standard model and the
 clean test it promises for the renormalizability of electroweak interaction
\cite{Hughes_98,Far_90}.

\section{Ground State Hyperfine Interval}

The ground state hyperfine splitting allows the most sensitive tests
of QED for the muon-electron interaction. An unambiguous and very precise atomic
physics value for the fine structure constant $\alpha$ can be derived from
$\Delta \nu _{HFS}$.
The experimental precision for $\Delta \nu _{HFS}$ has reached the level of
0.036ppm or 160Hz  which is just a factor of two above the estimated
contribution of -65Hz (15 ppb) from the weak interaction arising from an axial
vector -- axial vector coupling via Z-boson exchange \cite{Beg_74,Eid_96}. 
The sign of the weak effect is in muonium is opposite to the one for hydrogen,
because the positive muon is an antiparticle and the proton is a particle.
(Therefore the muonium atom may be viewed as a system which is neither
pure matter nor antimatter.)

With increased
experimental accuracy muonium can be the first atom where a shift in atomic
energy levels due to the weak interaction will be observed.
A contribution of 250 Hz (56 ppb) due to  strong interaction arises from
hadronic vacuum polarization.
The theoretical calculations for $\Delta \nu _{HFS}$
can be improved to the necessary
precision \cite{Kin_98}. The calculations themselves are at present accurate to about
50ppb. However, there is a 300ppb uncertainty arising from mass
$\mu_{\mu}$ of the positive muon.

\begin{table}[thb] \centering
\caption{Results extracted from the 1982 measurement of the 
muonium ground state hyperfine structure
splitting in comparison with some recent theoretical values and relevant quantities
from independent experiments.}

\begin{tabular}{llll}\hline
$\Delta \nu _{HFS}$(theory)     & 4463302.55 (1.33)(0.06)(0.18)kHz &(0.30ppm)&\cite{Eid_95}\\
$\Delta \nu _{HFS}$(theory)     & 4463303.27 (1.29)(0.06)(0.59)kHz &(0.35ppm)&\cite{Kar_96}\\
$\Delta \nu _{HFS}$(theory)     & 4463303.04 (1.34)(0.04)(0.16)(0.06)kHz &(0.30ppm)&\cite{Nio_97}\\
$\Delta \nu _{HFS}$(theory)     & 4463302.89 (1.33)(0.03)(0.00)kHz &(0.30ppm)&\cite{Sap_97}\\

$\Delta \nu _{HFS}$(expt.)      & 4463302.88(0.16)kHz    &(0.036ppm)&[7]\\
$\alpha ^{-1}$($\Delta \nu _{HFS}$)     & 137.035988(20)   &(0.15ppm)&\cite{Kin_97}\\
$\alpha ^{-1}$(electron g-2    )       & 137.035999 93(52)
&(0.004ppm)&\cite{Kin_97}\\
$\alpha ^{-1}$(condensed matter        )       & 137.0359979(32)  &(0.024ppm)&\cite{Cag_89}\\
$\mu_{\mu} / \mu_p$($\Delta \nu _{HFS}$)& 3.1833461(11)   &(0.36ppm)&[7]\\
$\mu_{\mu} / \mu_p$($\mu$SR in lq.bromine)& 3.1833441(17)   &(0.54ppm)&\cite{Kle_82}\\
\hline
\end{tabular}
\end{table}

All precision experiments to date have been carried out with muonium formed by
charge exchange after stopping $\mu^+$ in a suitable gas \cite{Hug_90}.
The experimental part of such an effort \cite{Hug_86} has been completed at
the Los Alamos Meson Physics Facility (LAMPF) and the recorded data are
presently being analyzed.
The experiment used a Kr gas target at typically
atmospheric pressures and a homogeneous magnetic field of about
1.7 Tesla. Microwave transitions between Zeeman levels which involve a muon spin flip
can be detected through a change in the spatial distribution of positrons from
muon decays, since due to parity violation in the decay process the positve muons decay
with the positrons preferentially emitted in muon spin direction.
The experiment  employed the technique of "old muonium" which allowed to reduce
the linewidth of the
signals can be reduced below half of the "natural" linewidth $\delta \nu_{nat}=
(\pi \cdot \tau_{\mu})^{-1}$=145kHz, where $\tau_{\mu}$ is the muon lifetime
of 2.2 $\mu$.
For this purpose the basically continuous beam of the LAMPF stopped muon channel
was chopped by an electrostatic kicker device into 4 $\mu$s long pulses with
14 $\mu$s separation.
Only atoms which were interacting
with the microwave field for periods longer than several muon lifetimes
were detected \cite{Bos_95}.
The basically statistically limited results improve
the knowledge of both zero field hyperfine splitting and muon magnetic moment
by a factor of three \cite{Kawall_98,Liu_98}.
The final results are discussed in detail in \cite{Kawall_98}. They
yield for the zero field splitting
\begin{equation}
\Delta \nu_{HFS}= 4 463 302 764(54) Hz ( 12 ppb)
\end{equation}
which agrees well with the most updated theoretical value \cite{Kin_98}
\begin{equation}
\Delta \nu_{theory}= 4 463 302 713(520)(34)(<100) Hz (120 ppb)~~ .
\end{equation}
For the magnetic moment one finds
\begin{equation}
\mu_{\mu}/\mu_p   = 3.183 345 26(39) (120 ppb)
\end{equation}
which translates into a muon-electron mass ratio of
\begin{equation}
m_{\mu}/m_e  = 206.768 270(24) (120 ppb).
\end{equation}

\begin{figure}[hbt]
\unitlength 1cm
 \begin{picture}(15,7.2)
\centering
  \epsfig{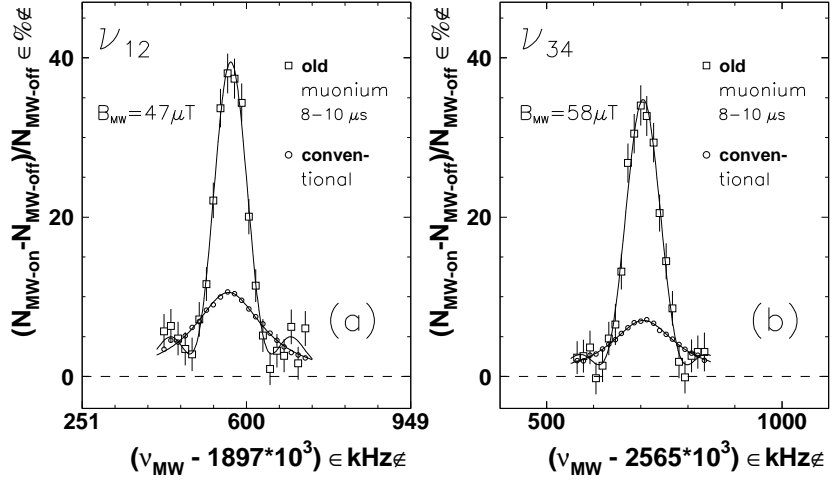}
 \end{picture}\par
 \caption[]{Conventional and 'old' muonium lines. The narrow 'old' lines are
 also higher. The frequencies $\nu_{12}$ and $\nu_{34}$ correspond to transitions
 between (a) the two energetically highest respectively (b) two lowest Zeeman sublevels
 of the n=1 ground state. As a consequence of the Breit-Rabi equation describing the
 behaviour of the levels in a magnetic field, the sum of these frequencies equals
 at any fixed field the zero field splitting $\Delta \nu$
 and the difference yields for a known
 field the muon magnetic moment $m_{\mu}$.}
\end{figure}

As the hyperfine splitting is proportional to
the fourth power of the fine structure constant $\alpha$, the improvement
in $\alpha$ will be much better and the value is comparable in
accuracy with the value of $\alpha$ determined using the Quantum Hall effect.
\begin{equation}
\label{alp}
\alpha^{-1}_{M-HFS}= 137.036 010 8(52) (39 ppb).
\end{equation} 
Here the value $\hq/m_e$ as determined with the help of measurements of
the neutron de Broglie wavelength \cite{Kru_97} 
has been benefitially employed
to gain higher accuracy compared to a previously used determination
based on the very well known Rydberg constant which yields
$\alpha^{-1}_{M-HFS}(traditional)= 137.035 998 6(80) (59 ppb)$. 
We can expect a near future small  improvement of the value in eq. (\ref{alp})
from ongoing determinations of $\hq/m_e$ through measurements of the 
photon recoil in Cs atom spectroscopy and a Cs atom mass measurement.

The limitation of $\alpha$  
from muonium HFS, when using eq. (\ref{alp}), arises mainly from to the muon mass.
Therefore any better determination of the muon mass,
respectively its magnetic moment,
e.g. through a precise measurement of the reduced mass shift in the muonium
1s-2s splitting, will result in an improvement of this value of $\alpha$.
It should be noted that already at present the good agreement within two
standard deviations between
the fine structure constant determined from muonium hyperfine structure
and the one from the electron magnetic anomaly is considered the
best test of internal consistency of QED, as one case involves bound state
QED and the other case QED of free particles.

\begin{figure}[hbt]
\unitlength 1cm
\centering
 \begin{picture}(10,8.24)
  \epsfig{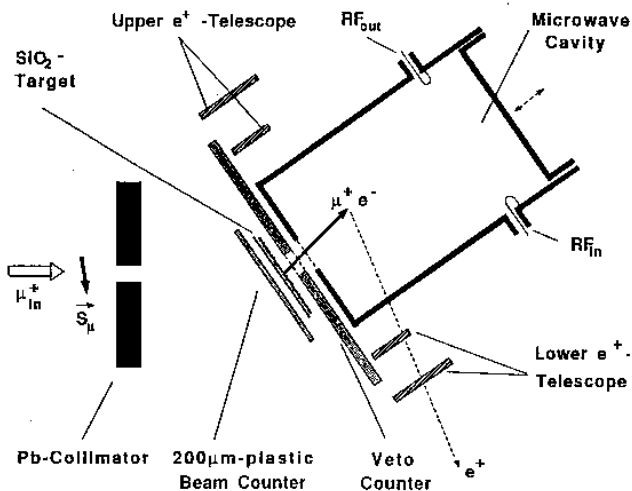}
 \end{picture}\par
 \caption[]{Principle of the setup for a measurement
of $\Delta \nu_{HFS}$ in vacuum. }
\end{figure}

\begin{figure}[bht]
\unitlength 1cm
\centering
 \begin{picture}(10,6.24)
  \epsfig{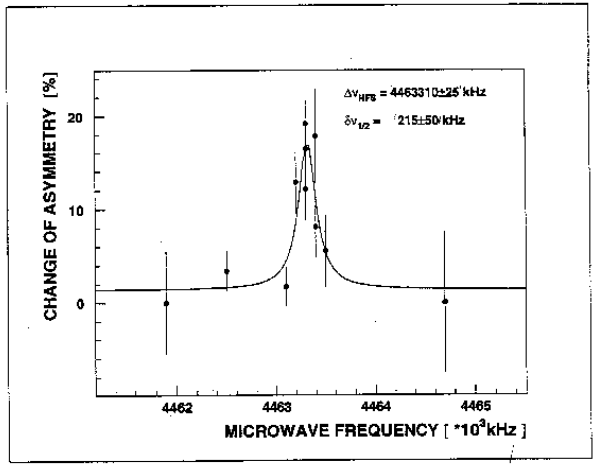}
 \end{picture}\par
 \caption[]{First observed muonium $\Delta \nu_{HFS}$ signal in vacuum.}
\end{figure}

The precision measurements of the muonium hyperfine splitting
performed so far, however,
suffer from the interaction of the muonium
atoms with the foreign gas atoms.
 With the discovery of polarized thermal muonium emerging from $SiO_2$ powder
targets into vacuum \cite{Woo_88} measurements of the $\Delta \nu _{HFS}$ are now
feasible in vacuum in the absence of a perturbing foreign gas.
Corrections for density effects are obsolete.
In a preliminary experiment at
the Paul Scherrer Institut (PSI) in Villigen, Switzerland,
the transitions between the $1^2S_{1/2},F=1$ and
the $1^2S_{1/2},F=0$ hyperfine levels could be induced in zero magnetic field
(see Fig. 1).
The atoms were formed
by electron capture after
stopping positive muons close to the surface of
a SiO$_2$ powder target. A fraction of these
diffused to the surface and left the
powder at thermal velocities (7.43(2)~mm/$\mu$s)
for the adjacent vacuum region.
A rectangular rf resonator operated in $TEM_{301}$
mode was placed directly above
the muonium production target. The atoms entered the cavity
with thermal velocities through a wall
opening. The $e^+$ from the parity violating muon decay $\mu^+
\rightarrow e^+ + \nu_e + {\bar \nu_{\mu}}$  were registered in two
scintillator telescopes which were mounted close to the side walls of the
resonator. The telescope axes have been oriented parallel respectively
antiparallel to the spin of the incoming muons which had been rotated transverse
to the muon propagation direction by an {\bf ExB} separator (Wien filter)
in the muon beam
line. At the resonance frequency of 4.46329(3)GHz a reduction of the muon
polarization of 16(2)\% has been observed as a signal from the hyperfine
transitions \cite{Jun_95}.

Certainly the method needs a lot of development work in order to improve
the signal strength. However, ultimately
one expects higher precision from experiments on muonium
in vacuum than from measurements in gases.

\section{Lamb shift in the first excited state}

The classical Lamb shift in the atomic hydrogen atom between
the metastable
$2^2S_{1/2}$ and the $2^2P_{1/2}$
states is totally QED in nature and seems to be ideal for
testing radiative QED corrections. An important contribution arises from the
internal proton structure. Its uncertainty limits any theoretical calculation.
Today, experiment and theory agree on the 10
ppm level. The purely leptonic muonium atom is free from nuclear structure
problems. In addition, the relativistic reduced mass and recoil contributions
are about one order of magnitude larger compared to hydrogen. Lamb shift
measurements at TRIUMF \cite{Oram_84} and LAMPF \cite{Bad_84,Ket_91}
have reached the $10^{-2}$
level of
precision. 

\begin{table}[thb] \centering
\caption{n=2 Lamb shift in muonium. Comparison between experiment and theory.}
\begin{tabular}{llll}\hline
Experiment&$\Delta \nu_{2^2S_{1/2}-2^2P_{1/2}}$[MHz]&experimental method &Ref.\\
\hline
TRIUMPF     &   1070(+12)(-15)  & direct $2^2S_{1/2}-2^2P_{1/2}$
trans.&\cite{Oram_84}\\
LAMPF       &   1042(+21)(-23)  & direct $2^2S_{1/2}-2^2P_{1/2}$
trans.&\cite{Bad_84}\\
LAMPF       &   1027(+30)(-35)  & extracted from
$2^2S_{1/2}-2^2P_{3/2}$ trans.&\cite{Ket_91}\\
Theory      &   1047.49(1)(9)&&\cite{Yen_92}\\
\hline
\end{tabular}
\end{table}

All these
measurements were carried out using fast muonium produced by a beam foil
method \cite{Bol_81} which is the only method known to date that produces
muonium in the metastable 2S state in usefull quantities.

\section{1s-2s Transition}

Doppler free excitation of the 1s-2s transition has been achieved in the past
at KEK in Tsukuba, Japan, \cite{Chu_88} and at the
Rutherford Appleton Laboratory (RAL) in Chilton, Didcot, UK \cite{Jun_91,Maas_94}.
The accuracy of the last experiment was limited by
ac Stark effect
and a frequency chirp caused by rapid changes of the
index of refraction in the dye solutions of the laser amplifier
employed in the laser system.
The experiment yielded $\Delta \nu_{{\rm 1S-2S}}$ = 2\,455\,529\,002 (33)(46)~MH z
 for the centroid 1S-2S transition frequency.
This value is
in agreement with theory
within two standard deviations \cite{Yen_92}.
The Lamb shift contribution to the 1S-2S
splitting has been extracted to $\Delta \nu_{{\rm LS}}$ = 6\,988 (33)(46)~MHz;
this the most precise
experimental Lamb shift value for muonium available today.
From the isotope shifts between
the muonium signal and the hydrogen and deuterium
1S-2S two-photon resonances we deduce the
mass of the positive muon as $m_{\mu}$ = 105.658\,80 (29)(43)~MeV/c$^2$.
An alternate  interpretation of the result yields the best test of the equality
of the absolute value of charge units in the first two generations of particles
at the $10^{-8}$ level \cite{Maas_94}.\\

A new measurement of the 1S-2S energy splitting of muonium
by Doppler-free two-photon spectroscopy has been performed
at
the worlds presently brightest pulsed surface muon source
which exists at RAL.
Increased accuracy is expected compared to a previously obtained value.
The series of experiments aims for an improvement of the muon mass.\\

\begin{figure}[hbt]
\unitlength 1cm
\centering
 \begin{picture}(11,5.24)
  \epsfig{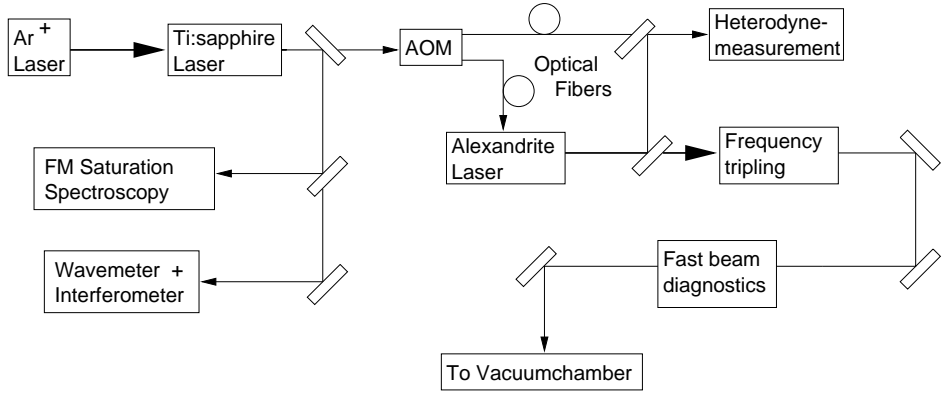}
 \end{picture}\par
 \caption[]{Laser system employed in the new muonium 1s-2s experiment.}
\end{figure}

The 1$^2$S$_{1/2}$(F=1) $\rightarrow$ 2$^2$S$_{1/2}$(F=1)
transition was induced by Doppler-free two-photon laser spectro\-sco\-py
using two counter-propagating laser beams of wavelength $\lambda = 244$~nm.
The necessary high power UV laser light
was generated
by frequency tripling the output of an alexandrite ring laser amplifier
in crystals of LBO and BBO.
Typically UV light pulses of energy 3~mJ and 80~nsec (FWHM) duration were used.
The alexandrite laser was seeded with light from a continuous wave
Ti:sapphire laser at 732~nm which was pumped by an Ar ion laser.
Fluctuations of the optical phase during the laser pulse were compensated with
an electro-optic device in the resonator of the ring amplifier to give
a frequency chirping of the laser light of less than about 5~MHz.
The laser frequency was calibrated by frequency modulation saturation
spectroscopy of a hyperfine component of the 5-13 R(26) line in thermally
excited iodine vapour.
The frequency of the reference line is about 700~MHz lower than 1/6 of the
muonium transition frequency.
The cw light was frequency up-shifted by passing through two
acousto-optic modulators (AOM's).
The muonium reference line has been calibrated preliminarily to 3.4~MHz at the
Institute of Laser Physics in Novosibirsk.
An independent calibration at the National Physics Laboratory (NPL) at
Teddington, UK, was performed to 140~kHz (0.35 ppb).

\begin{figure}[thb]
\label{m1s2s_sig}
\centering
\unitlength 1cm
 \begin{picture}(10,7.65)
  \epsfig{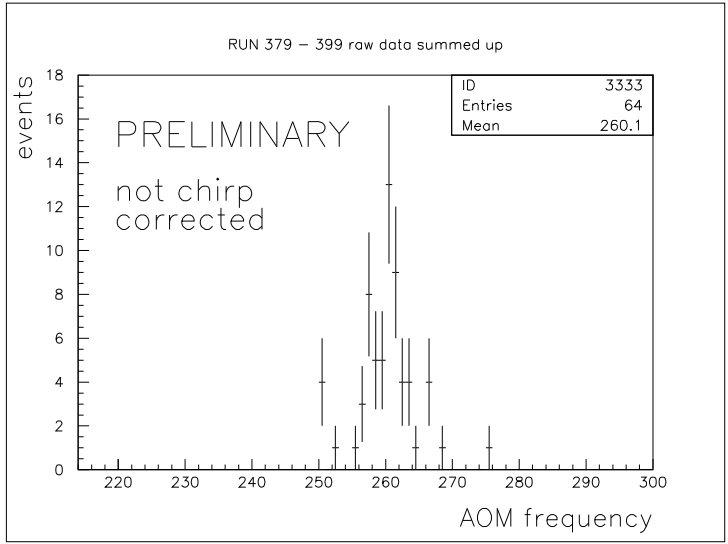}
 \end{picture}\par
 \caption[]{The
 1$^2$S$_{1/2}$(F=1)$\rightarrow$2$^2$S$_{1/2}$(F=1)
 transition signal in muonium, not corrected for sys\-te\-ma\-tic shifts due to
 frequency chirping and ac Stark effect.}
\end{figure}

The 1S-2S transition was detected by the
photoionization of the 2S state
by a third photon from the same laser field. The slow muon
set free
in the ionization process is accelerated to 2~keV and guided through a momentum
and energy selective path onto a microchannel plate particle detector (MCP).
Background due to scattered photons and other ionized particles
can be reduced to less than 1 event in 5 hours by shielding and
 by requiring that the MCP count falls into a 100~nsec wide window
centered at the expected time of flight for mu\-ons
and by additionally requiring
the observation of the energetic positrons from the muon decay.
On resonance an event rate of 9 per hour was observed.
The number of MCP events as a function of the laser frequency is displayed
in Fig. 6.

The line shape distortions due to frequency chir\-ping were investigated
theoretically using density matrix formalism to find
numerically the frequency dependence of the signal
for the time dependent laser intensity and  the
time dependent phase of the laser field which were both measured for
every individual laser shot using fast digitzed amplidude and
optical heterodyne signals \cite{Yak_96,Yak_98}.
The model was verified experimentally
at the ppb level by observing resonances in deuterium
and hydrogen. A careful analysis of the recorded muonium data is in progress
and promisses for the first succesful run with the new all solid state laser system
an accuracy of order 10~MHz which is significantly better than the previous result.
The measurements are still limited by properties of the laser system,
in particular by the chirp effect, rather than by statistics.

It can be expected that future experiments will reach well below  1 MHz
in accuracy promising an improved muon mass value.
The theoretical value at present is basically limited by the knowledge of recoil
terms at the 0.6 MHz level. Here some improvement of calculations will become important
for the next round of experiments.
\\

\section{Muonium to Antimuonium Conversion}

\begin{figure}[bht]
\unitlength 1.0 cm
 \begin{picture}(11.0,4.8)
  \centering
   \hspace{0.9 cm}
   \epsfig{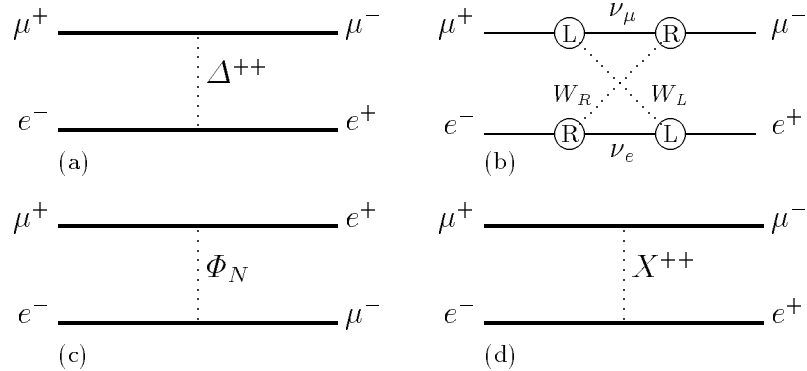}
 \end{picture}
 \centering\caption[]
        {\protect{\label{theo_mmb}}
        Muonium-antimuonium conversion in
        theories beyond the standard model. The interaction
        could be mediated by
        (a) a doubly charged Higgs boson $\Delta^{++}$
        \protect{\cite{Halprin_82,Herczeg_92}},
        (b) heavy Majorana neutrinos \protect{\cite{Halprin_82}},
        (c) a neutral scalar $\Phi_N$ \protect{\cite{Hou_96}}, e.g.
        a supersymmetric $\tau$-sneutrino $\tilde{\nu}_{\tau}$
        \protect{\cite{Halp_93,moha92}}, or
        (d) a dileptonic gauge boson $X^{++}$ \protect{\cite{Sasaki_94}}.
        }
\end{figure}

A spontaneous conversion of muonium (M=$\mu^+e^-$) into antimuonium
($\overline{\rm M=}\mu^-e^+$)
would violate additive lepton family
number conservation and is discussed  in many speculative
theories (see Fig. \ref{theo_mmb}).It would be an analogy in the lepton
sector to ${\rm K}^0$-$\overline{\rm K^0}$ oscillations \cite{Pon_58}.
Since lepton number is a solely empirical law and no underlying symmetry
could yet be revealed, it can be reliably applied only to the level
at which it has been tested. The M-$\overline{\rm M}$-conversion
process would in case of its existence, for example, cause a level splitting
of the muonium hyperfine states of order 519 Hz $\cdot {\rm G}_{{\rm M}\overline{\rm M}}$,
where $   {\rm G}_{{\rm M} \overline{\rm M}}$ is the coupling constant describing the
process in an effective four fermion coupling. Therefore the verification of
an upper limit for such an interaction is indispensable for a reliable extraction
of fundamental constants from spectroscopic measurements in muonium.

\protect{\label{mmbar}}

\begin{figure}[t]
\unitlength 1.0cm
\begin{minipage}{3.5cm}
              \caption[]{\protect{\label{mmbarsetup}}
              Top view of the MACS
              (Muonium - Antimuonium - Conversion - Spec\-trometer)
              apparatus at PSI to search for
              ${\rm M}-\overline{\rm M}$ - con\-vers\-ion
              \protect{\cite{Abela_96}}.  }
\end{minipage}
\hfill
\unitlength 1.0cm
\begin{minipage}{8.0cm}
\begin{picture}(5.5,6.7)
\hspace{0.0 cm}
\mbox{
\epsfig{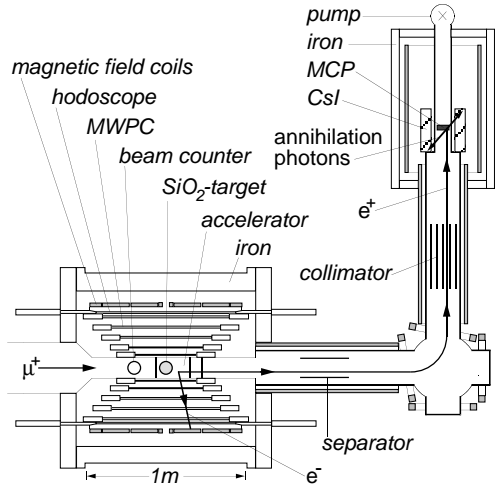}
        }
\end{picture}
\end{minipage}
\end{figure}

An experiment  set up at PSI
(Fig. \ref{mmbarsetup}) \cite{Jun_89,Abela_96} is designed to employ the signature
developed
in a predecessor experiment at LAMPF, which requires the coincident identification
of both particles forming the antiatom in its decay
\cite{Matthias_91,Willmann_97}.
Muonium atoms in vacuum with thermal velocities, which are produced
from a SiO$_2$ powder target, are observed for
antimuonium decays. Energetic electrons from the decay of
the $\mu^-$ in the antiatom can be observed in a magnetic spectrometer
at 0.1~T magnetic field
consisting of five concentric multiwire proportional chambers
and a 64 fold segmented hodoscope.
The positron in the atomic shell of the antiatom
is left behind after the decay with 13.5~eV average kinetic energy \cite{chat92}.
It can be accelerated to 7~keV in a two stage electrostatic device and guided
in a magnetic transport system onto a position sensitive microchannel
plate detector (MCP). Annihilation radiation can be observed in a 12 fold
segmented pure CsI calorimeter around it.

\begin{figure}[bht]
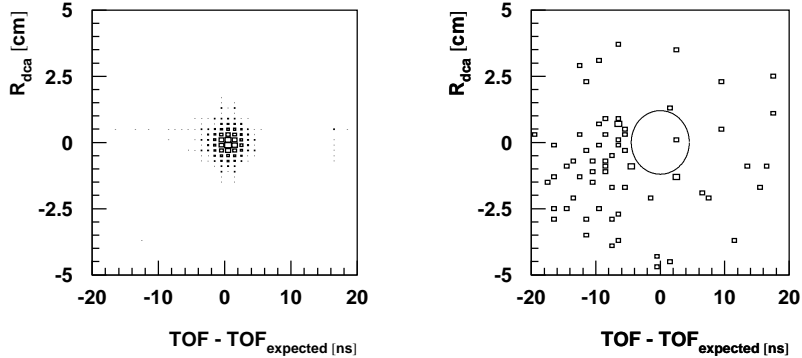

\unitlength 1.0 cm
 \begin{minipage}{4.7cm}
 \begin{picture}(5.0,4.7)
  \centering
   \hspace{-0.7 cm}
   \epsfig{figure=mmb_f3.ps,height=4.7cm}
 \end{picture}
 \end{minipage}
 \hspace{0.3cm}
 \begin{minipage}{4.7cm}
 \begin{picture}(5.0,4.7)
  \centering
   \epsfig{figure=mmb_f4.ps,height=4.7cm}
 \end{picture}
 \end{minipage}
 \centering\caption[] {\protect{\label{res_mmb} }
         Time of flight (TOF) and vertex quality for a muonium measurement
        (left) and the same for all data of the final 4 month search for antimuonium
        (right). One event falls into the indicated 3 standard deviations area.
        }
\end{figure}

The relevant measurements were performed during in total 6 month
distributed over 4 years during which $5.7 \cdot 10^{10}$ muonium atoms
were in the interaction region.
One event
fell within a 99\% confidence interval of all relevant distributions
(Fig. \ref{res_mmb}).
The expected background due to accidental coincidences is 1.7(2) events.
Thus an upper limit on the conversion
probability
of ${\rm P_{M\overline{M}} \leq 8.2\cdot 10^{-11}}/ {\rm S}_{\rm B}$
(90\% C.L.) was found,
where ${\rm S}_{\rm B}$ accounts for the interaction type dependent
suppression of the conversion
in the magnetic field of the detector  due to the removal of degeneracy between
corresponding levels in M and ${\rm \overline{M}}$.
The reduction is strongest for (V$\pm$A)$\times$(V$\pm$A),
where ${\rm S}_{\rm B}$=0.35 \cite{hori96,wong95}.
This yields for the traditionally quoted upper limit
on the coupling constant in effective four fermion interaction
 ${\rm G_{M\overline{M}}} \leq 3.0\cdot 10^{-3} {\rm G}_{\rm F} (90\%C.L.)$
with ${\rm G}_{\rm F}$ the weak interaction Fermi constant. \\

This new result, which exceeds bounds from previous experiments
\cite{Matthias_91,Gordeev_94}
by a factor of 2500 and the one from an early stage of the experiment
\cite{Abela_96}
by 35,
has some impact on speculative models.
A certain  ${\rm Z}_8$ model is ruled out with more than 4 generations of particles
where masses could be generated radiatively with heavy lepton seeding
\cite{Wong_94}.

A new lower limit of $m_{X^{\pm \pm}}$ $ \geq $ 2.6~TeV/c$^2$ $*g_{3l}$ (95\% C.L.)
on the masses of flavour diagonal bileptonic gauge bosons in GUT models
is extracted which lies well
beyond the value derived from direct searches, measurements of the muon
magnetic anomaly or high energy Bhabha scattering
\cite{Sasaki_94,Cuyp_96}. Here  $g_{3l}$ is of order 1 and depends on the details of the
underlying symmetry. For 331 models this translates into
$m_{X^{\pm \pm}}$ $ \geq $ 850~GeV/c$^2$ which excludes their minimal Higgs
version in which an upper bound of 600~GeV/c$^2$ has been extracted from an analysis of
electroweak parameters \cite{fram97,fram97a}. The 331 models
may still be viable in some extended form involving a Higgs octet \cite{fram98}.
In the framework of R-parity violating supersymmetry \cite{moha92,Halp_93}
the bound on the coupling parameters could be lowered by a factor of 15
to $\mid \lambda_{132}\lambda_{231}^*\mid \leq 3 * 10^{-4}$ for assumed
superpartner masses of 100 GeV/c$^2$.
Further the achieved level of sensitivity allows to narrow slightly the
interval of allowed
heavy muon neutrino masses
in minimal left-right symmetry \cite{Herczeg_92} (where a lower bound on
${\rm G_{M\overline{M}}}$ exists, if muon neutrinos are heavier than 35 keV)
to $\approx$ 40~keV/c$^2$ up to the present
experimental bound at 170~keV/c$^2$.

In minimal left right symmetric models, in which ${\rm {M\overline{M}}}$ conversion is allowed,
the process is intimately connected to
the lepton family number violating muon decay
$\mu^+ \rightarrow e^+ + \nu_{\mu} + \overline{\nu}_e$.
With the limit achieved in this experiment
this decay is not an option
for explaining the excess neutrino counts in the LSND neutrino
experiment at Los Alamos  \cite{herc97,LSND_96}.

The consequences for atomic physics of muonium are such that the expected level
splitting in the ground state due to ${{\rm M} - \overline{{\rm M}}}$
interaction is below $1.5~{\rm Hz} / \sqrt{S_B}$ reassuring  the validity of
 fundamental constants determined in muonium spectroscopy.

A future  ${{\rm M} - \overline{{\rm M}}}$  experiment could take particularly
advantage of high intense pulsed beams.
In contrast to other lepton flavour violating muon decays,
the conversion through its nature
as particle - antiparticle oscillation,
has a time evolution in which the probability for finding
$\overline{{\rm M}}$ in the ensemble remaining after muon decay
increases quadratically in time, giving the signal an advantage growing in time
over major exponentially decaying background \cite{Willmann_97}.

\section{Long term future possibilities}

It appears that the availability of particles limits the ability to
increase noticeably the accuracy of spectroscopy experiments on muonium.
Therefore any measure to boost the  particle fluxes would be a very important
step forward.
In principle, we need significantly more intense accelerators, such
as they are presently discussed at various places.
In the intermediate future the Japanese Hadron Facility (JHF) or a possible
European Spallation Source (ESS) are important options.
Also the discussed Oak Ridge neutron spallation source
could in principle accommodate intense muon beams.
The most promising facility would be, however, a muon collider \cite{Palmer_98},
the front end of which could provide
muon rates 5-6 orders of
magnitude higher than present beams (see Table \ref{muon_fluxes}).

%
\begin{table}[bht]
 \caption[]{\protect{\label{muon_fluxes}} Muon fluxes of
 some existing and future facilities, Rutherford Appleton Lab\-orat\-ory (RAL),
 Japanese Hadron Facility (JHF), European Spallation Source (ESS), Muon collider (MC). }
\begin{tabular}{|c|cccccc|}
\hline
                    &RAL($\mu^+$)    &PSI($\mu^+$)   & PSI($\mu^-$)  &JHF($\mu^+$)$^\dag$
                    &ESS($\mu^+$)     &MC ($\mu^+$, $\mu^-$)\\
 \hline
 Intensity ($\mu$/s)& $3\times 10^6$ &$3\times 10^8$ &$1\times 10^8$ &$4.5\times 10^7$
                     &$4.5\times 10^7$ &$7.5\times 10^{13}$ \\
 Momentum bite   &&&&&&\\
 \hspace*{4mm} $\Delta$ pm/p[\%] & 10& 10            & 10            & 10
  & 10              & 5-10     \\
  Spot size     &&&&&&\\
 (cm $\times$ cm)         & 1.2$\times$2.0 &3.3$\times$2.0  &3.3$\times$2.0  &
 1.5$\times$2.0 &1.5$\times$2.0 & few$\times$few \\
 Pulse structure    & 82 ns & 50 MHz    & 50 MHz    & 300 ns & 300 ns & 50 ps\\
                    & 50 Hz & continuous & continuous&  50 Hz &  50 Hz & 15 Hz\\
\hline
  \end{tabular}
  {\footnotesize
  $^\dag$ Recent studies indicate that the $10^{11}$ particles/s region might be reachable
  \cite{Kuno_98}.}
\end{table}

At short term, at PSI the muon beam of the $\pi$E3 area could be chopped to give
pulses of typically 1 $\mu$s duration and about 10 $\mu$s separation with in total beam
rates of up to $5 \cdot 10^{6} \mu^+$/s, which would be 5 times higher rate than
at LAMPF when the recently completed muonium hyperfine structure experiment was performed.
\\

With such improvements in the particle fluxes one could expect increased accuracy for
the muonium hyperfine splitting and the muon magnetic moment. For the 1s-2s splitting
a cw laser experiment could just about become feasible as well as many other
important laser experiments \cite{Bos_96,Kaw_98}
With the large number of very interesting possibilities
for measuring
fundamental constants and testing of basic physical laws muonium spectroscopy appears
to be as lively as is now for almost three decades.

\acknowledgments{ It is a pleasure to thank the organizers of the
                  workshop for creating a wonderful and
                  stimulating atmosphere and for their great hospitality. 
                  This work was in part supported by the German BMBF and a NATO
                  research grant.}

\end{document}